\documentclass{aastex62}

\received{}
\revised{}
\accepted{}
\submitjournal{}

\shorttitle{Measuring the absolute total intrinsic redshifts of the main sequence stars and red giants}
\shortauthors{Dai et al.}

\begin{document}

\title{Measuring the absolute total intrinsic redshifts (surface gravity plus the convective blueshift) of the main sequence stars and red giants using GAIA data}

\correspondingauthor{De-Chang Dai, ZhiGang Li}
\email{diedachung@gmail.com, zhigli@sjtu.edu.cn}

\author{De-Chang Dai}
\affiliation{Center for Gravity and Cosmology, School of Physics Science and Technology, \\
Yangzhou University, 180 Siwangting Road, Yangzhou City, Jiangsu Province, P.R. China 225002
}
\affiliation{
CERCA/Department of Physics/ISO, Case Western Reserve University, Cleveland OH 44106-7079
}

\author{ZhiGang Li}
\affiliation{College of Physics and Electronic Engineering, Nanyang Normal University, Nanyang, Henan, 473061, China}
\affiliation{Department of astronomy, Shanghai Jiao Tong University, Shanghai 200240, China}

\author{Dejan Stojkovic}
\affiliation{HEPCOS, Department of Physics, SUNY at Buffalo, Buffalo, NY 14260-1500}



\begin{abstract}

We analyze the GAIA release II data to demonstrate how one can measure the absolute total intrinsic redshifts of the main sequence stars and red giants.
We remove the relative velocity components of the stars' motion with respect to the sun by doing the analysis in the local standard of rest (LSR) frame defined by the average stars' motion. We provide results for four different types of stars. F, G and K types of stars have about the same value of intrinsic redshift, which is however much smaller than the expected gravitational redshift. This indicates that the GAIA's data include convective blueshift effect of a several hundreds m/s magnitude. The red giants' intrinsic redshifts are negative, which implies that their convective blueshift is stronger than the gravitational redshift. This is expected since red giants are far less compact than other types.

\end{abstract}


\section{Introduction}
Various techniques utilized in observational astrophysics enabled us to measure general relativistic effect such as gravitational redshift of distant stars.
The first confirmed measurement of gravitational redshift of a distant star comes from the measurement of the apparent radial velocity of a while dwarf, Sirius B \citep{1925Obs....48..337A}. The sun's gravitational redshift has also been measured (see a review in \citep{2012SoPh..281..551T}). Recently the gravitational redshift was observed from the star S2 orbiting around the massive black hole candidate SgrA* \citep{2018A&A...615L..15G}. Apart from the sun, which is the closest, and white dwarfs which have very strong gravitational redshift, it is possible to measure gravitational redshift in a group of comoving stars by comparing average redshifts of different types of stars and removing the (Doppler) velocity redshift. This method has a shortcoming that it provides only a relative rather than absolute gravitational redshift. In addition, it was found in \citep{2011A&A...526A.127P} that the M67 open cluster does not produce an expected signal. The discrepancy can be caused by the convective blueshift which is wavelength dependent \citep{2017A&A...597A..52M,2017A&A...607A.124M}. It is also possible to measure radial velocities using astrometric methods without involving spectroscopic data\citep{1999A&A...348.1040D}. The intrinsic redshift can be extracted by comparing astrometric and spectroscopic radial velocities. Based on this method, Le{\~a}o et al studied the Hyades open cluster and found that red giants have more blueshifted spectra than the dwarfs\citep{2018arXiv181108771L}. However, there are very few data sets that include both spectroscopic and astrometric radial velocities, so methods based on analyzing only spectroscopic radial velocities data are still very useful.

In this paper we demonstrate how one can extract absolute total intrinsic redshifts of the main sequence stars and red giants. This intrinsic redshift is usually dominated by the surface gravity of the star (which is purely a general relativistic effect), however it also includes the convective blueshift which cannot be easily removed without additional information. A straightforward way to measure the total redshift effect is to observe a star at rest with respect to us. This is practically impossible, due to the stars relative motion with respect to the sun. However, we can circumvent the problem by using a local standard of rest (LSR) frame as the coordinate system of average stars' motion \citep{2010MNRAS.403.1829S,2011MNRAS.412.1237C,2015MNRAS.449..162H}. Then, we can use these coordinates to study the gravitational redshift. Such procedure was previously utilized in \citep{2010ApJ...712..585F} to analyze gravitational redshift of 449 non-binary white dwarfs. Since gravitational redshift of white dwarfs is $2-3$ orders of magnitude stronger than that of the main sequence stars, several hundred stars are enough for the procedure to work. For the same method to be useful for the main sequence stars, this number must be much higher.

\section{data and method}

\begin{figure}
   \centering
\includegraphics[width=10cm]{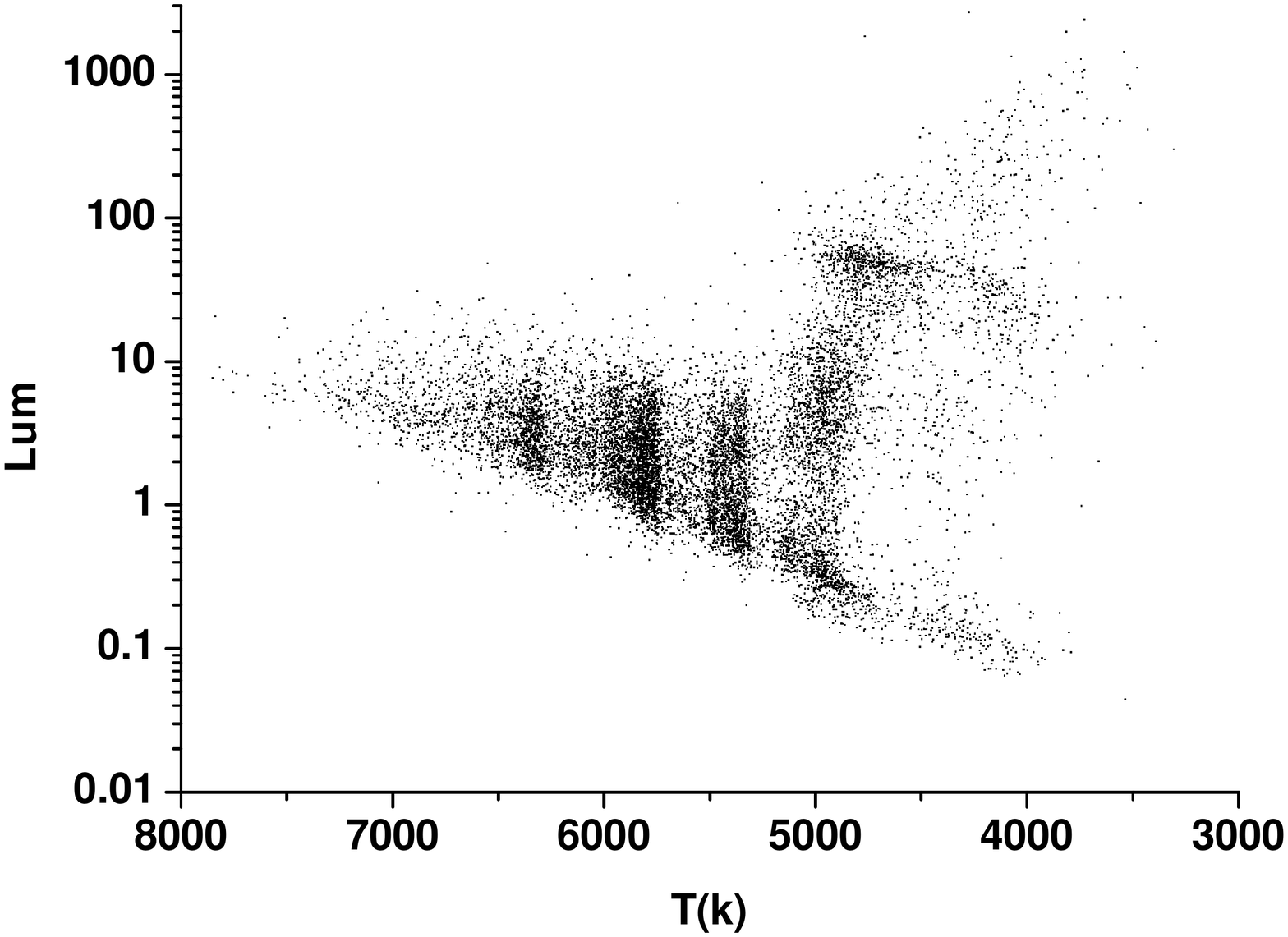}
\caption{This HR diagram includes stars whose distance to the sun is from $0.05$kpc to $1$kpc. Only a small fraction (1/200) of stars we use are drawn in this diagram.
}
\label{Hr}
\end{figure}

In this work we use the GAIA data release II. GAIA sky survey charts a three-dimensional map of our Galaxy, the Milky Way, revealing the composition, formation and evolution of the Galaxy\citep{2016A&A...595A...1G,2018A&A...616A...1G}. The star morphology can be obtained from the HR-diagrams in \citep{2018A&A...616A..10G,2018A&A...616A..11G}. In our study, we concentrate on the kinematics information of the nearby stars. The HR-diagram for stars within $1$kpc from the sun is shown in fig. \ref{Hr}. Most of the stars' surface temperatures range are from $3500$k to $8000$k. This covers the F, G, K, and the red giant types, and therefore we will mainly focus on these four groups.

GAIA data provide the total redshift of the stars. This means that both the gravitational and Doppler redshift effects are included. To distinguish between these two effects, the redshift velocity $V_r$\citep{2018arXiv180409369C,2018arXiv180409372K,2018A&A...616A...7S} and the transverse velocity, $V_{Ra}$ and $V_{Dec}$ (in the right ascension and declination direction), must be considered simultaneously. $V_{Ra}$ and $V_{Dec}$ are measured relative to the sun. $V_r$ is measured from the star's spectral redshift, which does not directly represent the relative velocity to the sun along the line of sight direction. $V_r$ includes the Doppler redshift due to the star's motion, gravitational redshift, and convective blueshift due to the motion of the material close to the star's surface. We can separate $V_r$ in two components as
\begin{equation}
V_r= v_r + v_s.
\end{equation}
Here, $v_r$ is the star's relative velocity with respect to the sun along the line of sight direction, while $v_s$ is the combination of the star's gravitational redshift, $v_g$, and convective blueshift, $v_c$. The $v_g$ component is about the same for the same type of stars, since the star's mass and radius do not change much within the class. However, $v_c$ can be strongly dependent on the observation wavelength \citep{2017A&A...597A..52M,2017A&A...607A.124M}. It was found in \citep{2017A&A...597A..52M,2017A&A...607A.124M} that  the convective blueshift can be several hundreds of m/s in the GAIA's wavelength region.  Many additional effects can in principle affect the results. For example, gravitational potential of the galaxy can affect the radial velocity measurements \citep{2003A&A...401.1185L}. However, since our stars are within $1$kpc, the redshift from the Milky way's gravitational potential is only several tens of m/s. This is one order of magnitude smaller than $v_r$, so we neglect it in the study. Relativistic effects due to the difference between the proper and coordinate time (as defined by an adopted metric) can also affect the radial velocity measurement. However, this effect is about several m/s and can be safely neglected in our study\citep{2018arXiv181108771L,2003A&A...401.1185L}. Stellar rotation and activity can also affect the spectral radial velocity measurement. The magnitude of the stellar rotation effect is expected to be about several $10m/s$ in \citep{2018arXiv181108771L}, though Lindegren \& Dravins expect a value of several $100m/s$\citep{2003A&A...401.1185L}. Since it is very difficult to have such a detailed information about every single star, we just assume that these effects are randomly distributed and therefore statistically canceled out. Finally, the effect of stellar activity is expected to be several $10m/s$\citep{2018arXiv181108771L,2003A&A...401.1185L}. It appears that the surface gravitational redshift and convection blueshift are the dominant effects, while the other effects are smaller than our precision, so we primarily focus on them. 

\begin{figure}
   \centering
\includegraphics[width=10cm]{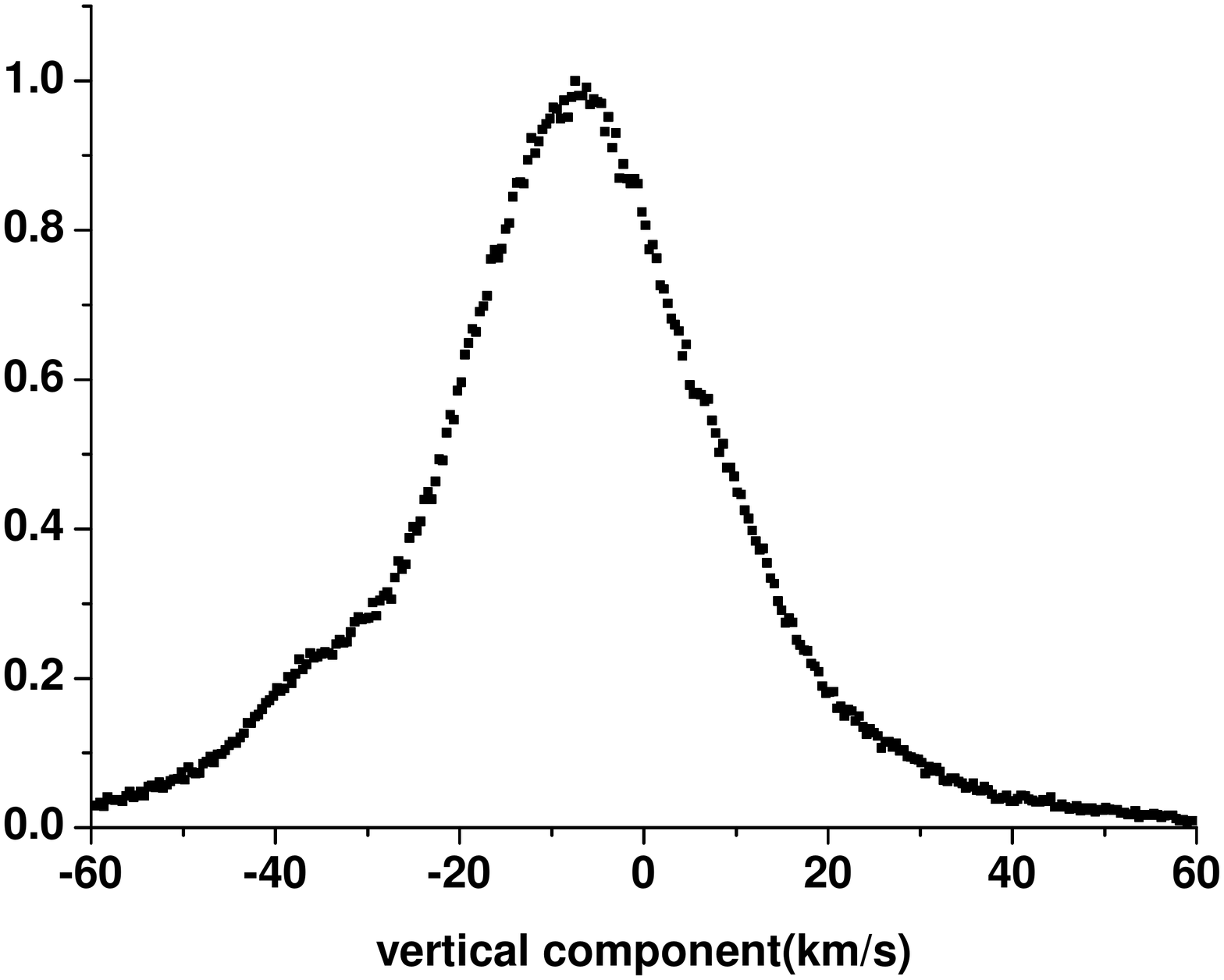}
\caption{The distribution of the stars' vertical velocity components (Eq.~\ref{vertical}). The distances between the stars and the sun range from $0.05$kpc to $0.5$kpc. Only stars with temperatures between $5200$K to $6000$K (G-type stars) and radius between $0.8$ to $1.2$ are included. The distribution is normalized to a maximum magnitude of $1$.
}
\label{W}
\end{figure}

Our main goal is to extract $v_s$ from $V_r$, $V_{Ra}$ and $V_{Dec}$. It is possible to remove the relative velocity components by doing the analysis in the local standard of rest frame.  A star's velocity in this frame can be labeled as $(U, V, W)$, where $U$, $V$ and $W$ are velocities in the galactic radial, rotational, and vertical direction respectively. Accordingly, the sun's velocity in the same frame is $(U_\odot, V_\odot, W_\odot)$. In the local standard of rest frame, all the stars are moving in a group, and the average values of all three components, $\bar{U}$, $\bar{V}$ and $\bar{W}$, should not be far from $0$. Unfortunately, this is not completely correct. The $U$ and $V$ components are more complicated \citep{2010MNRAS.403.1829S,2014CeMDA.118..399F}, since $U$ is not completely symmetric \citep{2010MNRAS.403.1829S}, and the galaxy is expanding in radial direction \citep{2011MNRAS.412.2026S}. Thus, the contribution of $U$ and $V$ on measuring $v_s$ are not easily estimated. $W$ on the other hand is expected to be symmetric, since Milky Way is basically cylindrically symmetric. We therefore analyze the stars velocities in the vertical direction.  The vertical component is then
\begin{equation}
(V_r \hat{r}+V_{Dec} \hat{\delta}+V_{Ra}\hat{\alpha})\cdot \hat{z}=W-W_\odot +v_s \cos(b) .
\label{vertical}
\end{equation}
Here, $\delta$ and $\alpha$ are declination and right ascension in the equatorial coordinate system respectively. $\hat{r}$, $\hat{\delta}$ and $\hat{\alpha}$ are the unit vectors in three relevant directions. $\hat{z}$ is the galactic north pole direction. The declination and right ascension of the galactic north pole's direction are $\alpha_g=192.85948^\circ$ and $\delta_g=27.12825^\circ$. $b$ is the galactic latitude.
The right hand side of equation \ref{vertical} is theoretical expectation, while the left hand side is what we fit.
Since the values of $W$ are randomly distributed, they can be treated as noise, while the relevant information to be extracted is in quantities $v_s$ and $W_\odot$. Fig. \ref{W} shows the left hand side of equation \ref{vertical}. The distribution is practically spherically symmetric, except some small bumps.

 The data are separated into four different groups corresponding to different types of stars: F, G, K, and red giants. F type includes star surface temperatures from $6000$K to $7500$k, and star radii from $1.15R_\odot$ to $1.4R_\odot$. We use the total of $134697$ F type stars in our analysis. G type includes star surface temperatures from $5200$K to $6000$k, and star radii from $0.8R_\odot$ to $1.2R_\odot$. We use the total of $361157$ G type stars in our analysis. K type includes stars surface temperatures from $3700$K to $5200$k, and star radii from $0.7R_\odot$ to $1R_\odot$.  We use the total of $137456$ K type stars in our analysis. Red giant type includes stars surface temperatures from $3700$K to $5200$k, and star radii larger than $10R_\odot$.  We use the total of $155721$ red giants in our analysis. All the stars are between $0.05$pc to $1$kpc away from the sun. Since  a main sequence star's surface gravitational redshift is around several hundreds of m/s, and the vertical velocity component is about several tens of km/s, we need tens of thousands of stars to achieve the signal to noise ratio around one. We therefore include several hundred thousands of stars of each type in our analysis, which should be enough to study the star's intrinsic redshift effect.

We did not include all the stars within $1$kpc from the sun, because the stars' radii are not always given very precisely. Fortunately, the catalog gives the stars' radii for the main sequence stars and red giants, which is sufficient for the main aim of our study. We choose only stars with radii larger than $10R_\odot$ as red giants, in order to avoid overlap with other large stars. We discard all of the other stars. The number of stars we included within these four types is quite sufficient to distinguish the signal from the noise.

\section{Result}
\begin{figure}
   \centering
\includegraphics[width=10cm]{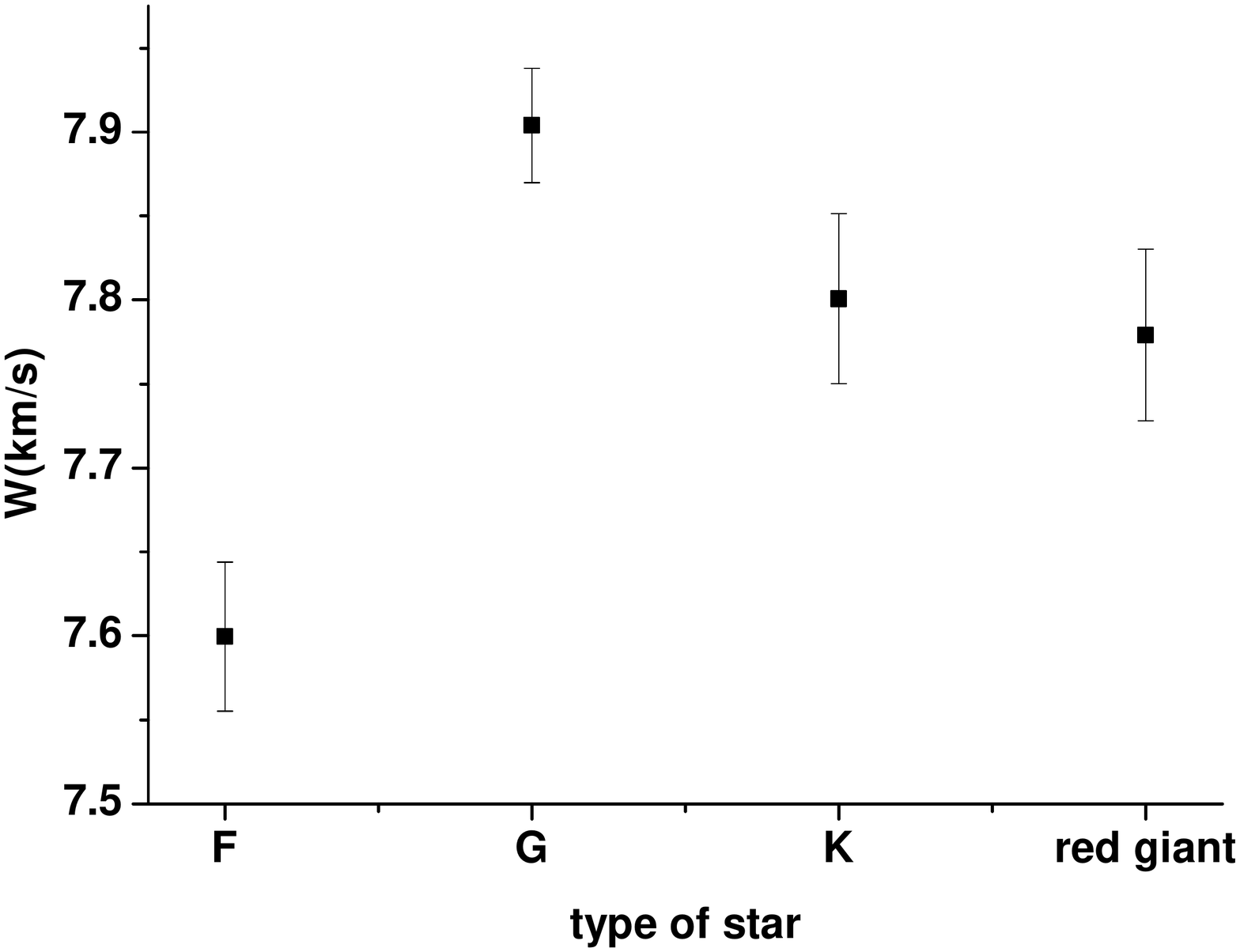}
\caption{The sun's vertical velocity, $W_\odot$, with respect to the LSR. The error bars are $1\sigma$. The values for $W_\odot$ are quite consistent for the four different types of stars (F, G, K, and red giants).
}
\label{vertical-w}
\end{figure}

We fit the vertical data with Eq.~(\ref{vertical}), using the standard minimal $\chi^2$ analysis. Both $W_\odot$ and a star's redshift can be obtained at the same time. The error for each data point is assumed to be the same, and can be estimated from vertical component's standard deviation, which is about several tens of km/s. Fig. \ref{vertical-w} shows $W_\odot$ from different type of stars. The values for $W_\odot$ for F type stars are a bit lower than the values for other types. The apparent difference might come from systematic errors, or from the stars' spatial distribution \citep{2010MNRAS.403.1829S,2011MNRAS.412.1237C,2015MNRAS.449..162H}. However, this is still within $3 \sigma$, so it should not matter much. In the literature, the measured values for $W_\odot$ range from $3.6$km/s to $10$km/s \citep{2015MNRAS.449..162H}. The result depends on the samples and models used in analysis. With our fitting strategy, the local standard of rest's velocity should be close to the average value of stars' vertical motion.  Since different types of stars cover different regions of space, this may cause the above mentioned discrepancy.

\begin{figure}
   \centering
\includegraphics[width=10cm]{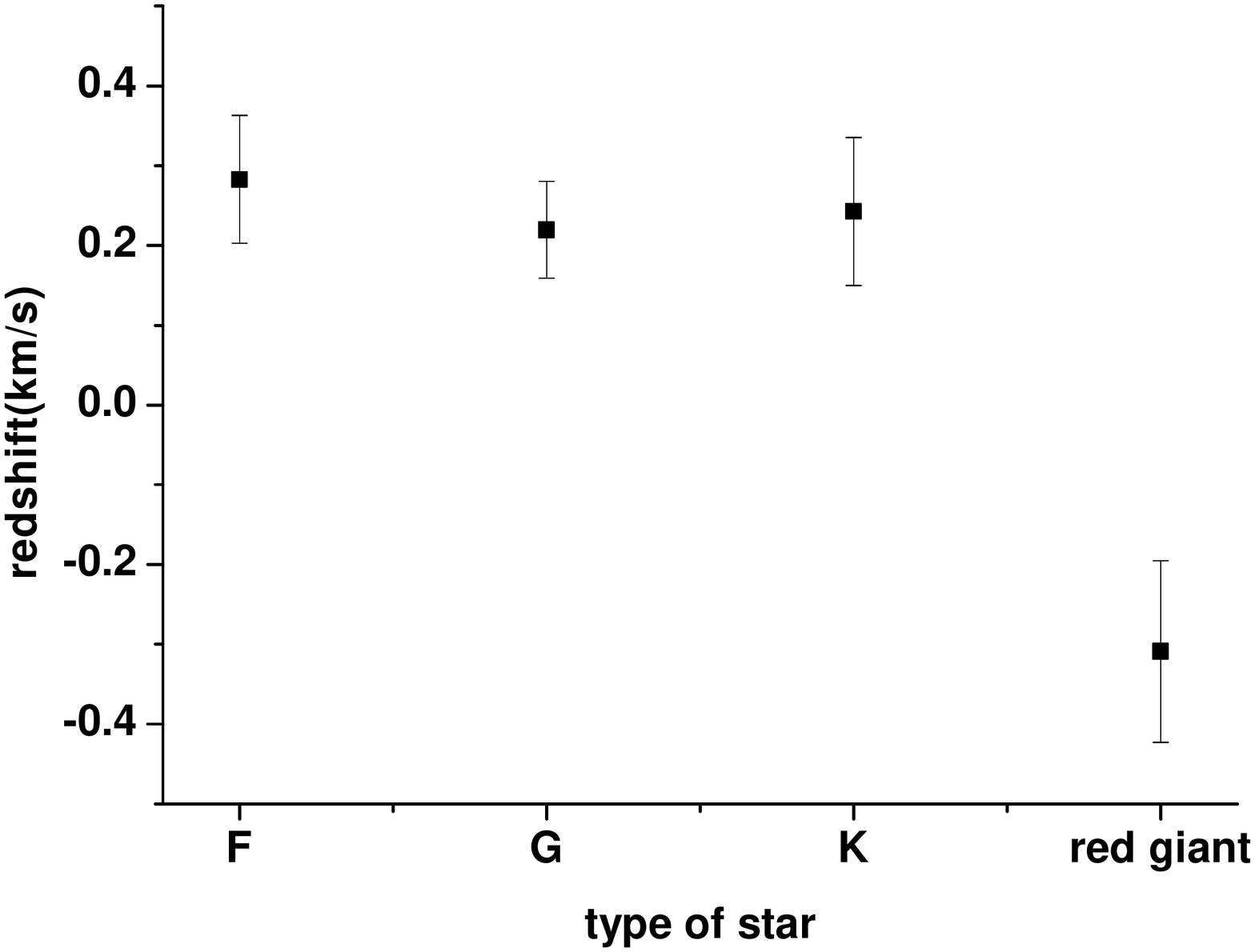}
\caption{We plot here the total intrinsic redshift (gravitational plus convective blueshift) represented by the velocity $v_s$ for different types of stars. The error bars are $1\sigma$.  The red giants' redshift values are negative, which implies that their convective blueshift is stronger than the gravitational redshift. F,G and K types have about the same value of $v_s$, which is larger than the red giants value. The gravitational redshift difference between the sun and a $10$ solar radii giant is about $570$ m/s. The difference can be larger for an even larger red giant. The difference in $v_s$ values between G-type stars and red giants is $\Delta v_s =0.53\pm 0.13$ km/s, which is consistent with the expected difference coming from gravitational redshift ($\approx 600$m/s), though the exact convection blueshift contribution is not known here.
}
\label{redshift}
\end{figure}

Fig. \ref{redshift} shows $v_s$ (the total gravitational and convective blueshift) from different types of stars. F, G and K have about the same $v_s$. This value, however, is much smaller than the expected gravitational redshift, which is about $600$ m/s. This may confirm that the GAIA's data does indeed include convective blueshift effect of a several hundreds m/s magnitude\citep{2017A&A...597A..52M,2017A&A...607A.124M}. The red giants $v_s$ are much smaller than those of the main sequence stars. This is consistent with the theoretical prediction. Since red giants $v_s$ are negative, their convective blueshift, $v_c$, must be stronger than its gravitational redshift, $v_g$. This is also expected since red giants radii are much larger, and therefore $v_g$ is expected to be smaller.  The difference in $v_s$ values between G-type stars and red giants is $\Delta v_s =0.53\pm 0.13$ km/s, which is consistent with the expected difference coming from gravitational redshift ($\approx 600$m/s). We note however that in this study we do not know the exact convection blueshift contribution for different types of stars.

\section{conclusion}

In conclusion, we presented here the first measurement of the absolute total intrinsic redshifts (which include surface gravity plus the convective blueshift) of the main sequence stars and red giants using GAIA data. The obtained values are consistent with the theoretical expectations. As a next step, it would be very important to separate the surface gravity from the convective blueshift effect. If mass and radius of a star are known, then one can calculate the gravitational potential and extract the net convective blueshift effect. This can perhaps be done in binary systems.

\acknowledgments

D.C Dai was supported by the National Science Foundation of China (Grant No. 11433001 and 11775140), National Basic Research Program of China (973 Program 2015CB857001) and  the Program of Shanghai Academic/Technology Research Leader under Grant No. 16XD1401600. ZL was supported by the Project for New faculty of Shanghai Jiao Tong University (AF0720053), the National Science Foundation of China (No. 11533006, 11433001) and the National Basic Research Program of China (973 Program 2015CB857000). DS was partially supported by the NSF grant PHY 1820738.

This work has made use of data from the European Space Agency (ESA) mission
{\it Gaia} (\url{https://www.cosmos.esa.int/gaia}), processed by the {\it Gaia}
Data Processing and Analysis Consortium (DPAC,
\url{https://www.cosmos.esa.int/web/gaia/dpac/consortium}). Funding for the DPAC
has been provided by national institutions, in particular the institutions
participating in the {\it Gaia} Multilateral Agreement.


\begin{thebibliography}{}

\bibitem[Adams(1925)]{1925Obs....48..337A} Adams, W.~S.\ 1925, The Observatory, 48, 337

\bibitem[Co{\c s}kuno{\v g}lu et al.(2011)]{2011MNRAS.412.1237C} Co{\c s}kuno{\v g}lu, B., Ak, S., Bilir, S., et al.\ 2011, \mnras, 412, 1237

\bibitem[Cropper et al.(2018)]{2018arXiv180409369C} Cropper, M., Katz, D., Sartoretti, P., et al.\ 2018, arXiv:1804.09369


\bibitem[Dravins et al.(1999)]{1999A&A...348.1040D} Dravins, D., Lindegren, L., \& Madsen, S.\ 1999, \aap, 348, 1040

\bibitem[Falcon et al.(2010)]{2010ApJ...712..585F} Falcon, R.~E., Winget, D.~E., Montgomery, M.~H., \& Williams, K.~A.\ 2010, \apj, 712, 585

 \bibitem[Francis \& Anderson(2014)]{2014CeMDA.118..399F} Francis, C., \& Anderson, E.\ 2014, Celestial Mechanics and Dynamical Astronomy, 118, 399

\bibitem[Gaia Collaboration et al.(2016)]{2016A&A...595A...1G} Gaia Collaboration, Prusti, T., de Bruijne, J.~H.~J., et al.\ 2016, Astronomy \& Astrophysics, 595, A1

\bibitem[Gaia Collaboration et al.(2018)]{2018A&A...616A...1G} Gaia Collaboration, Brown, A.~G.~A., Vallenari, A., et al.\ 2018, Astronomy \& Astrophysics, 616, A1

\bibitem[Gaia Collaboration et al.(2018)]{2018A&A...616A..10G} Gaia Collaboration, Babusiaux, C., van Leeuwen, F., et al.\ 2018, Astronomy \& Astrophysics, 616, A10

\bibitem[Gaia Collaboration et al.(2018)]{2018A&A...616A..11G} Gaia Collaboration, Katz, D., Antoja, T., et al.\ 2018, Astronomy \& Astrophysics, 616, A11

\bibitem[Gravity Collaboration et al.(2018)]{2018A&A...615L..15G} Gravity Collaboration, Abuter, R., Amorim, A., et al.\ 2018, Astronomy \& Astrophysics, 615, L15

\bibitem[Huang et al.(2015)]{2015MNRAS.449..162H} Huang, Y., Liu, X.-W., Yuan, H.-B., et al.\ 2015, mnras, 449, 162

\bibitem[Katz et al.(2018)]{2018arXiv180409372K} Katz, D., Sartoretti, P., Cropper, M., et al.\ 2018, arXiv:1804.09372

 \bibitem[Le{\~a}o et al.(2018)]{2018arXiv181108771L} Le{\~a}o, I.~C., Pasquini, L., Ludwig, H.-G., \& de Medeiros, J.~R.\ 2018, arXiv:1811.08771
 
 \bibitem[Lindegren \& Dravins(2003)]{2003A&A...401.1185L} Lindegren, L., \& Dravins, D.\ 2003, aap, 401, 1185
 
\bibitem[Meunier et al.(2017a)]{2017A&A...597A..52M} Meunier, N., Lagrange, A.-M., Mbemba Kabuiku, L., et al.\ 2017a, Astronomy \& Astrophysics, 597, A52

 \bibitem[Meunier et al.(2017b)]{2017A&A...607A.124M} Meunier, N., Mignon, L., \& Lagrange, A.-M.\ 2017b, Astronomy \& Astrophysics, 607, A124
 
\bibitem[Pasquini et al.(2011)]{2011A&A...526A.127P} Pasquini, L., Melo, C., Chavero, C., et al.\ 2011, Astronomy \& Astrophysics, 526, A127

\bibitem[Sch{\"o}nrich et al.(2010)]{2010MNRAS.403.1829S} Sch{\"o}nrich, R., Binney, J., \& Dehnen, W.\ 2010, mnras, 403, 1829

\bibitem[Siebert et al.(2011)]{2011MNRAS.412.2026S} Siebert, A., Famaey, B., Minchev, I., et al.\ 2011, mnras, 412, 2026

\bibitem[Soubiran et al.(2018)]{2018A&A...616A...7S} Soubiran, C., Jasniewicz, G., Chemin, L., et al.\ 2018, Astronomy \& Astrophysics, 616, A7

\bibitem[Takeda \& Ueno(2012)]{2012SoPh..281..551T} Takeda, Y., \& Ueno, S.\ 2012, Solar Physics, 281, 551

\end{thebibliography}
\end{document}